\documentclass[a4paper,fleqn]{cas-dc}

\usepackage[numbers]{natbib}

\usepackage{color}

\newcommand{\red}{\color{red}}

\def\tsc#1{\csdef{#1}{\textsc{\lowercase{#1}}\xspace}}
\tsc{WGM}
\tsc{QE}

\begin{document}
\let\WriteBookmarks\relax
\def\floatpagepagefraction{1}
\def\textpagefraction{.001}

\shorttitle{$\pi$-solitons on a ring}    

\shortauthors{Ivanov, Kartashov}  

\title [mode = title]{$\pi$-solitons on a ring of waveguides}  



%

\author[1]{Sergey K. Ivanov}[orcid=0000-0001-9230-1035]

\cortext[1]{Corresponding author}



\ead{sergey.ivanov@icfo.eu}



\affiliation[1]{organization={ICFO-Institut de Ciencies Fotoniques, The Barcelona Institute of Science and Technology},
            addressline={Av. Carl Friedrich Gauss, 3}, 
            postcode={08860},
            city={Castelldefels (Barcelona)},
            country={Spain}}



\author[2]{Yaroslav V. Kartashov}[orcid=0000-0001-8692-982X]




\affiliation[2]{organization={Institute of Spectroscopy, Russian Academy of Sciences},
            addressline={Fizicheskaya Str., 5}, 
            city={Troitsk, Moscow},
            postcode={108840}, 
            country={Russia}}




\begin{abstract}
We study the existence and stability of $\pi$-solitons on a ring of periodically oscillating waveguides. The array is arranged into Su–Schrieffer–Heeger structure placed on a ring, with additional spacing between two ends of the array. Due to longitudinal oscillations of waveguides, this Floquet structure spends half of the longitudinal period in topological phase, while on the other half it is nontopological. Nevertheless, waveguide oscillations lead to the emergence of anomalous topological $\pi$-modes at both ends of the structure that strongly couple in our ring geometry, leading to the formation of previously unexplored in-phase and out-of-phase $\pi$-modes. We study topological solitons bifurcating from such linear $\pi$-modes and demonstrate how their properties and stability depend on the size of the ring and on spacing between two ends of the array.
\end{abstract}




\maketitle


The past decade has witnessed growing interest to photonic topological insulators representing extension of topological materials originally discovered in solid-state physics to optical realm offering rich prospects for implementation of topologically protected states in practical switching and routing devices, and lasers. Various proposals for observing photonic topological insulators have been made~\cite{Lu-14,Ozawa-19}. Among them is the realization of photonic Floquet topological insulator in a waveguiding system~\cite{Rechtsman-13}. The Floquet protocol, resulting from periodic modulation of waveguiding system in the direction of light propagation~\cite{Rudner-20,Bandres-16,Biesenthal-22}, has been used to demonstrate anomalous topological phases~\cite{Mukherjee-17,Maczewsky-17}, unpaired Dirac cones~\cite{Leykam-16(2)}, tunneling inhibition~\cite{Zhang-18(1)} and coupling of edge states~\cite{Zhang-18(2),Zhong-19}, etc. Remarkably, such modulations can also create anomalous Floquet topological $\pi$-modes, whose localization strongly depends on parameters of modulation~\cite{Asboth-14,Lago-15,Fruchart-16,Zhang-17,Petracek-20}. One of the simplest systems allowing observation of these localized periodically oscillating anomalous Floquet states is the Su-Schrieffer-Heeger (SSH) array with dynamically modulated intra- and inter-cell coupling strengths~\cite{SuSchrieffer-79}. Experimentally topological $\pi$-modes were observed only in linear regime in plasmonic and acoustic systems~\cite{Cheng-19,Wu-21,Sidorenko-22,Cheng-22}, as well as in waveguide arrays~\cite{Song-21}.

On the other hand, photonic systems may possess strong nonlinear response that in combination with nontrivial topology of the material leads to many intriguing phenomena~\cite{Smirnova-20,Maczewsky-20}. Among them is the formation of topological edge solitons, existing in diverse forms and geometries. Such states have been predicted and observed in helical waveguide arrays~\cite{Lumer-13,Leykam-16(1),Ablowitz-14,Mukherjee-21,Ivanov-20(1),Ivanov-20(2),Ivanov-21}, in topological systems admitting linearly coupled co-propagating edge states~\cite{Zhang-19,Smirnova-19,Ivanov-20}, and in second-order insulators ~\cite{Kirsch-21,Hu-21}, to name just a few systems. Straight SSH arrays and their generalizations also allow formation of topological edge solitons~\cite{Solnyshkov-17,Xia-20,Bongiovanni-21,Kartashov-22}. At the same time, the impact of nonlinearity on anomalous Floquet systems remains practically unexplored. It is only very recently that $\pi$-mode solitons have been predicted in the simplest modulated line and square SSH waveguide arrays~\cite{Zhong-23}.


The goal of the present Letter is to introduce anomalous $\pi$-modes in new configuration, where modulated SSH-like waveguide array is placed on a ring, such that coupling of states at the opposite ends of the same structure becomes possible. We show that localized anomalous $\pi$-modes emerging at both ends of ring SSH array due to periodic oscillations of waveguides couple to form more complex in-phase and out-of-phase oscillating states with quasi-propagation constants in topological gap. In focusing medium such states give rise to in-phase and out-of-phase $\pi$-solitons that can be simultaneously stable. We study how the properties of such solitons depend on the size of the ring and on spacing between two ends of the SSH ring array.


\begin{figure*}[t!]
\centering
\includegraphics[width=\linewidth]{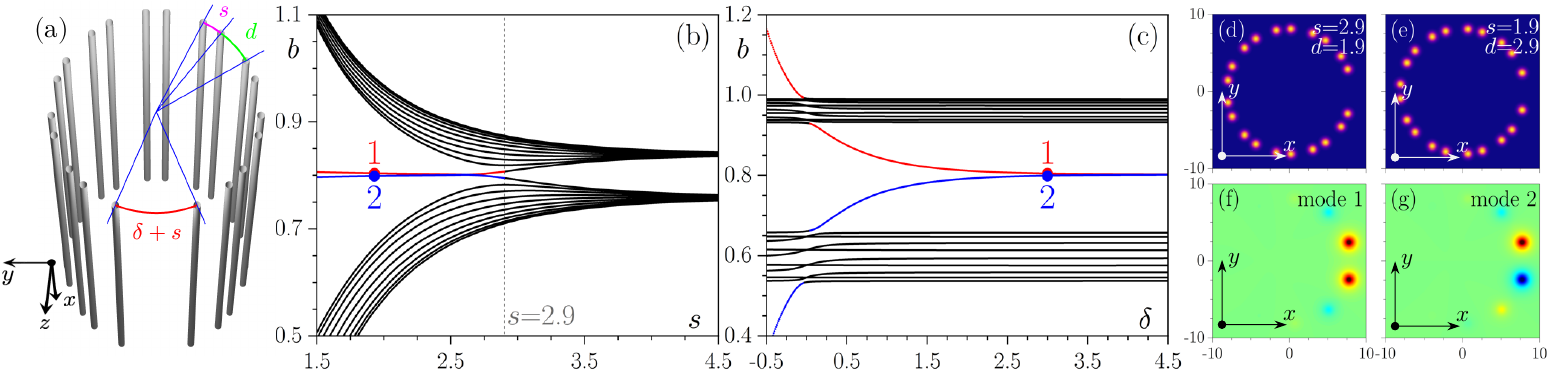}
\caption{
(a) Example of the SSH ring array with straight waveguides. (b) Spectrum of modes of the array with $N=10$ dimers versus inter-cell spacing $s$ for fixed $d=2.9$ and $\delta=3$. Vertical dashed line indicates the transition from trivial to topological phase. (c) Eigenvalues of modes supported by the topological array as functions of spacing $\delta$ for $s=1.9$, $d=2.9$. Array profiles for $s=2.9$, $d=1.9$ (d) and $s=1.9$, $d=2.9$ (e). Examples of the modes corresponding to the red (f) and blue (g) dots in panels (b), (c). Here and below $p=5$ and $\sigma=0.5$.
}
\label{fig1}
\end{figure*}

\begin{figure*}[t!]
\centering
\includegraphics[width=\linewidth]{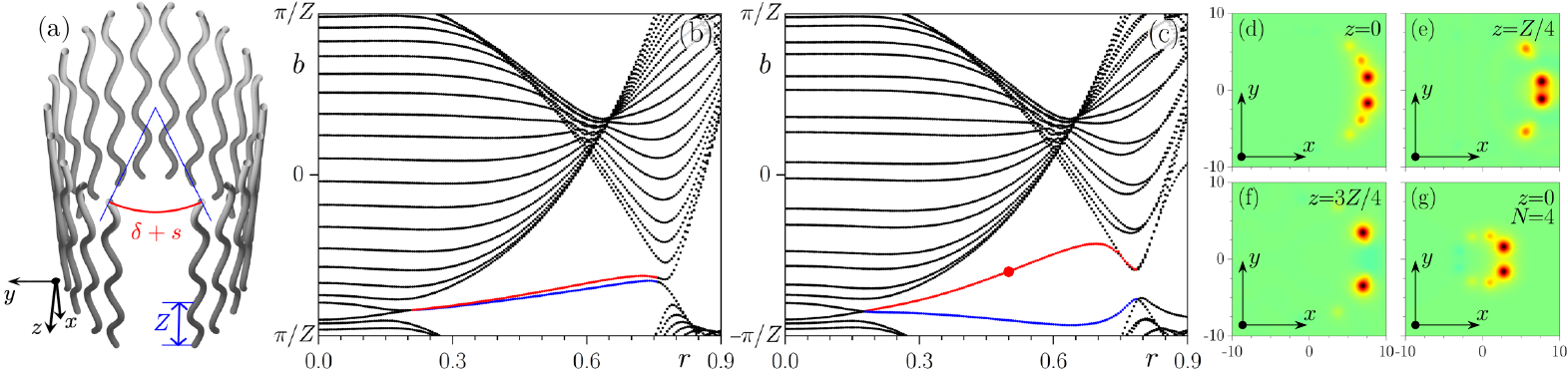}
\caption{
(a) Example of the modulated SSH ring array. Floquet spectra of the modulated array with $\delta=3$ (b) and $\delta=1$ (c) for various $r$ values. Examples of $\pi$-modes corresponding to the red dot in panel (c) at different distances $z$ within one longitudinal period for the array with $N=10$ (d)-(f) and for the array with $N=4$ at $z=0$ (g). Here and below $s=d=2.4$ and $Z=17$.
}
\label{fig2}
\end{figure*}

We consider paraxial propagation of light in a modulated waveguide array with focusing nonlinearity, which can be described by the dimensionless nonlinear Schr\"{o}dinger equation for the field amplitude $\psi$
\begin{equation} 
\label{NLS}
    i \frac{\partial \psi}{\partial z} =  -\frac{1}{2} \nabla^2 \psi  - \mathcal{R}(x,y,z)\psi - |\psi|^2\psi.
\end{equation}
Here $\nabla=(\partial/\partial x,\partial/\partial y)$,
the $x$ and $y$ coordinates are normalized to the characteristic transverse scale $r_0=10$~$\mu \textrm{m}$, the propagation distance $z$ is normalized to the diffraction length $kr_0^2$, $k = 2\pi n_0/\lambda$, $\lambda=800$~$\textrm{nm}$, and $n_0\approx 1.45$ is the unperturbed refractive index of the material. Dimensionless intensity $|\psi|^2=1$ corresponds to peak intensity of $I=n_0|\psi|^2/k^2r_0^2n_2\approx 8\cdot10^{15} \, \textrm{W}/\textrm{m}^2$, where the effective nonlinear refractive index is $n_2\sim1.4\cdot10^{-20} \, \textrm{m}^2/\textrm{W}$.
The function $\mathcal{R}(x,y,z) = p\sum_{m=0}^{N-1} [Q_{m1}(x,y,z)+Q_{m2}(x,y,z)]$ describes $2N$ single-mode Gaussian waveguides placed on a ring of radius $\rho=(Ns+Nd+\delta)/2\pi$, where $Q_{mi} = e^{-[(x-\rho \cos\phi_{mi}(z))^2+\left(y-\rho \sin\phi_{mi}(z)\right)^2]/\sigma^2}$, $i=1,2$, $\phi_{m1}(z)=[(2m+1)s/2+md+\delta/2-r\sin(\omega z)]/\rho$ and $\phi_{m2}(z)=[(2m+1)s/2+(m+1)d+\delta/2+r\sin(\omega z)]/\rho$, $s$ and $d$ are the inter-cell and intra-cell arc spacings, respectively, $\delta$ is the additional spacing between two ends of the array, $\omega = 2\pi/Z$, $Z$ is the longitudinal modulation period, and $r$ is the amplitude of waveguide oscillations. Each guide has a width of $\sigma = 0.5$. The depth of the waveguides is given by $p = k^2r_0^2\delta n/n_0$, where $\delta n$ is the refractive index contrast. Here we set $p = 5$ which corresponds to $\delta n\approx5.6\cdot10^{-4}$. Schematic illustrations of static $(r=0)$ and modulated $(r\ne0)$ ring SSH arrays are presented in Fig. \ref{fig1}(a) and \ref{fig2}(a), respectively.

By neglecting the nonlinear term in Eq.~(\ref{NLS}) and introducing the ansatz $\psi = u(x,y,z)\exp(ibz)$, where $b$ represents a quasi-propagation constant and $u(x, y, z)=u(x, y, z+Z)$ is a complex field of $z$-periodic Floquet mode, we obtain the problem $bu = (\partial^2 u/\partial x^2 + \partial^2 u/\partial y^2)/2 + Ru + i \partial u/\partial z$ that can be solved numerically using propagation and projection method~\cite{Leykam-16(2),Ivanov-20(1)}. First, it is instructive to consider spectrum of static ring array [Fig.~\ref{fig1}(a)] with straight waveguides (in this case $r = 0$ and $b$ is a standard propagation constant), presented in Figs.~\ref{fig1}(b) and \ref{fig1}(c) for $N = 10$ dimers. For this array, when $s$ becomes smaller than $d$ [see Fig.~\ref{fig1}(e)] the inter-cell coupling exceeds the intra-cell one and array enters topological phase, when two topological edge states emerge between two bulk bands. These states, depicted by red and blue lines in Fig.~\ref{fig1}(b), represent in-phase and out-of-phase combinations of localized modes emerging at the opposite ends of the structure [see modes $1$ and $2$ in Figs.~\ref{fig1}(f) and~(g)]. Their localization becomes more pronounced when $s$ decreases, while the intra-cell spacing  $d=2.9$ and $\delta=3$ remain fixed. The width of the topological gap, where edge states appear, expands as $s$ decreases. When $s > d$ [see Fig.~\ref{fig1}(d)], the array is in trivial phase, and no edge states can form within trivial gap in the spectrum. In Fig.~\ref{fig1}(c) we show how linear spectrum depends on the additional spacing $\delta$ for $s=1.9$ and $d=2.9$. Notice that topological modes appear only for $\delta>0$, and the difference of their propagation constants decreases with increase of $\delta$, since coupling between two ends of array becomes weaker.

Next we consider periodically modulated ring array illustrated in Fig.~\ref{fig2}(a). This array is in the ``instantaneous'' topological phase on half of the $Z$ period, but on the other half it is in the trivial phase, where edge states would not exist in static structure. Nevertheless, for a certain range of amplitudes of waveguide oscillations $r$, the modulated array supports topological $Z$-periodic anomalous $\pi$-modes emerging at two ``coupled'' ends of the array [see red and blue lines in Floquet spectra of such structure in Figs.~\ref{fig2}(b) and~(c)]. It can be seen that $\pi$-modes (red and blue lines) emanate at $r\approx0.2$ from the point of overlap of the bulk band (black lines) with its own Floquet replica arising due to periodicity of the spectrum in the vertical direction. Their localization increases with $r$. Importantly, $\pi$-modes corresponding to red and blue lines are Floquet generalizations of in-phase and out-of-phase edge states in ring configuration. Their quasi-propagation constants merge when spacing $\delta$ increases and two ends of array stop interacting, while when $\delta\rightarrow 0$ they separate more and more and tend to merge with bulk bands. Anomalous $\pi$-modes undergo significant shape variations during propagation, while remaining exactly $Z$-periodic. This behavior is evident in Figs.~\ref{fig2}(d)--(f) for $N=10$, where we display Re$[\psi]$ distributions in such modes at selected distances for $r = 0.5$. Although the intensity distributions are identical at distances $z = 0, Z/2, Z$, the profiles at $z = Z/4$ and $z = 3Z/4$ differ, highlighting the fact that the full period corresponds to $Z$. We also show the example of $\pi$-mode at $z=0$ for smaller ring with only $N=4$ dimers in Fig.~\ref{fig2}(g).

\begin{figure}[t!]
\centering
\includegraphics[width=\linewidth]{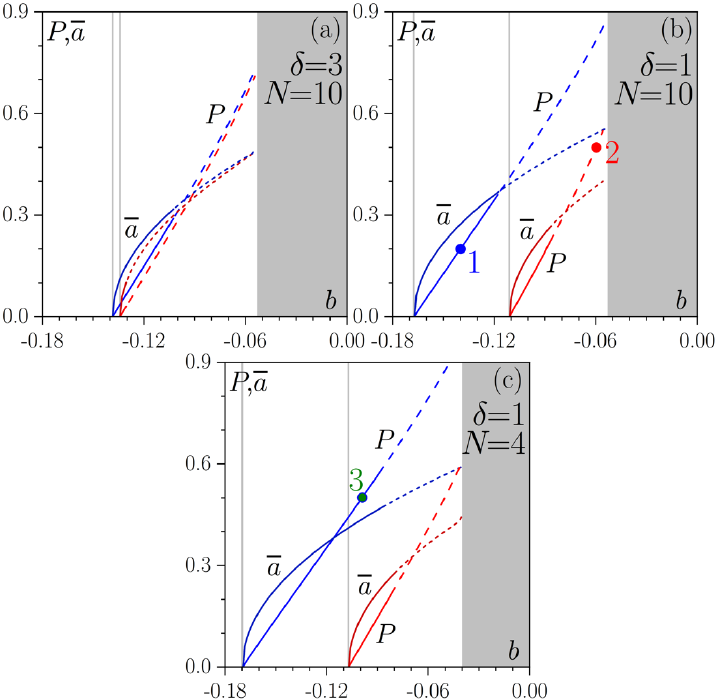}
\caption{
Averaged peak amplitude $\overline{a}$ and power $P$ versus quasi-propagation constant $b$ for $\pi$-soliton families at $r=0.5$ in the arrays with $N=10$, $\delta=3$ (a) and $\delta=1$ (b) and in the array with $N=4$, $\delta=1$ (c). Dashed lines represent unstable solitons, solid lines represent stable ones, and the shaded regions show bulk bands.
}
\label{fig3}
\end{figure}


We now turn to $\pi$-solitons, which bifurcate from linear anomalous $\pi$-modes in this ring configuration. Fig.~\ref{fig3} illustrates the dependence of power $P=\iint |\psi|^2dxdy$ of $\pi$-soliton, as well as of its averaged peak amplitude $\overline{a} = Z^{-1} \int a\, dz$, where $a = {\rm max}|\psi|$, on quasi-propagation constant $b$ for $r = 0.5$. Here, red and blue lines correspond to in-phase and out-of-phase $\pi$-solitons that can co-exist in topological gap. The power of such states $P$ increases almost linearly away from bifurcation point from corresponding linear modes (shown by vertical gray lines). Nonlinearity allows to adjust the localization of $\pi$-solitons by changing location of $b$ within the gap. At sufficiently high powers, when $b$ approaches the bulk band (shaded regions), the solitons interact with bulk modes and develop long tails in the entire ring.

\begin{figure}[t!]
\centering
\includegraphics[width=\linewidth]{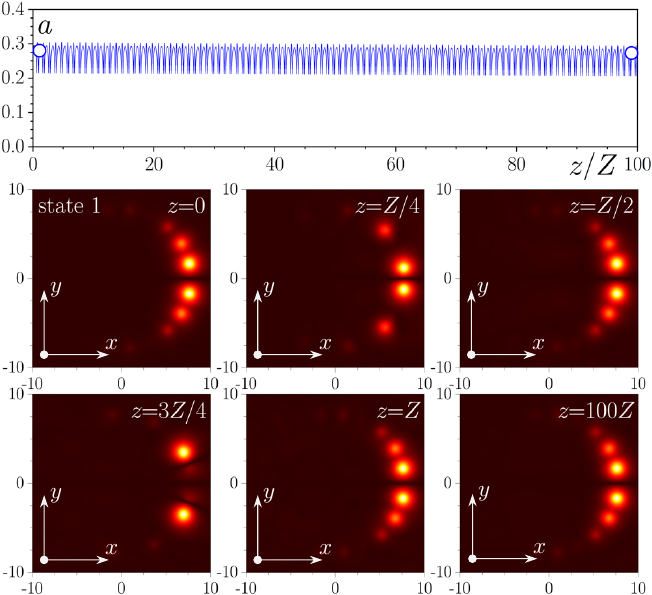}
\caption{
Stable propagation of the out-of-phase $\pi$-soliton. Top row shows peak amplitude $a$ vs normalized distance $z/Z$, while middle and bottom rows show field modulus distributions at different $z$ corresponding to the dots in the top panel. Soliton corresponds to the blue dot in Fig. 3(b).
}
\label{fig4}
\end{figure}

\begin{figure}[t!]
\centering
\includegraphics[width=\linewidth]{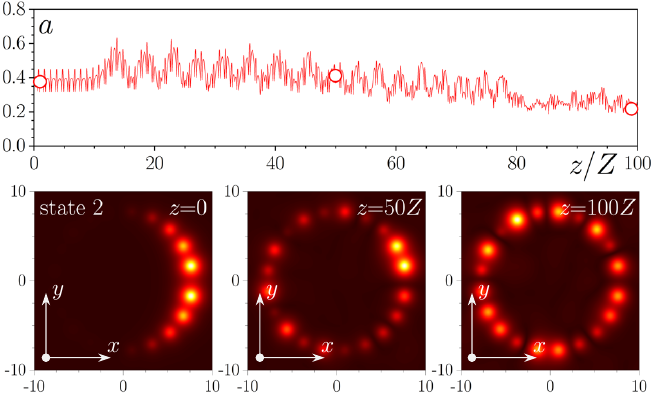}
\caption{
Decay of the unstable in-phase $\pi$-soliton. Top row shows peak amplitude $a$ vs normalized distance $z/Z$, while bottom row shows field modulus distributions at different $z$. This soliton corresponds to the red dot in Fig. 3(b).
}
\label{fig5}
\end{figure}

In Fig.~\ref{fig3}, solid branches correspond to stable $\pi$-solitons, while dashed branches correspond to unstable ones. Stability was analyzed by modeling the propagation of perturbed solitons $\psi|_{z=0} = u\left(x,y,z=0\right)\left(1 + \delta_{{\rm re}} + i\delta_{{\rm im}}\right)$, where $\delta_{{\rm re}} + i\delta_{{\rm im}}$ is a small complex noise with amplitude uniformly distributed within the segment $[-0.05, +0.05]$. At $N=10$ and $\delta=3$ [Fig.~\ref{fig3}(a)], solitons become unstable when their power exceeds $P\approx0.03$ for in-phase states (red line) and $P\approx0.25$ for out-of-phase states (blue line). Interestingly, for smaller spacing $\delta=1$ [Fig.~\ref{fig3}(b)] stability regions expand, so that in-phase solitons become stable up to $P\approx0.23$, while out-of-phase ones up to $P\approx0.37$. An example of stable propagation of the out-of-phase $\pi$-soliton with peak intensity $\sim5\cdot10^{14}\, \textrm{W}/\textrm{m}^2$ is shown in Fig.~\ref{fig4}, while Fig.~\ref{fig5} shows typical instability development for in-phase $\pi$-soliton. In the former case, the soliton shows persistent oscillations over hundreds of periods periodically recovering its shape despite considerable transformations that it experiences inside each period $Z$ (see field modulus distributions in Fig. \ref{fig4}). In contrast, the unstable $\pi$-soliton radiates and eventually decays, expanding over the entire ring (see Fig. \ref{fig5}).

\begin{figure}[t!]
\centering
\includegraphics[width=\linewidth]{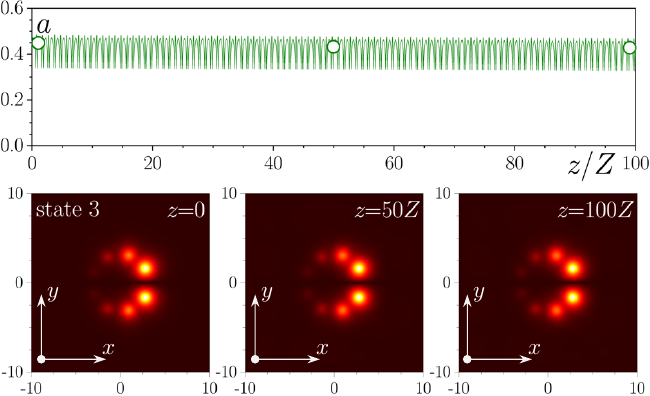}
\caption{
Stable propagation of the out-of-phase $\pi$-soliton in array with $N=4$. This soliton corresponds to the green dot in Fig. 3(c).
}
\label{fig6}
\end{figure}

By decreasing the size of the array to $N=4$ dimers, one can dramatically expand the domains of stability for $\pi$-solitons, as depicted in Fig.~\ref{fig3}(c). This illustrates a direct influence of the size of the ring on stability of such states. The propagation dynamics of stable soliton with peak intensity $\sim1.3\cdot10^{15}\, \textrm{W}/\textrm{m}^2$ for this small-size ring array is shown in Fig.~\ref{fig6}.
We also investigated the influence of the oscillation amplitude $r$ on the stability of $\pi$-solitons. Our observations indicate that the stability intervals progressively expand as the value of $r$ increases. However, it is important to note that this trend holds true only up to an upper limit of $\sim 0.7$ for $r$. Beyond this threshold, the introduction of radiation becomes a significant factor, affecting the stability of the solitons.

Summarizing, we have shown that ring SSH arrays composed from oscillating waveguides can support two different types of stable $\pi$-solitons with different phase structure, emerging due to coupling of two ends of the same array in ring geometry. Stability of such states strongly depends on the size of ring array and on spacing between two ends of the structure. These results show that anomalous Floquet systems may be used as a platform for observation of many intriguing nonlinear phenomena.

\medskip


\section*{Funding}

ICFO was supported by Agencia Estatal de Investigación (CEX2019-000910-S, PGC2018-097035-B-I00); Departament de Recerca i Universitats de la Generalitat de Catalunya (2021 SGR 01448); CERCA; Fundació Cellex; and Fundació MirPuig. Y.V.K.’s academic research was supported by Russian Science Foundation (grant 21-12-00096) and research project FFUU-2021-0003 of the Institute of Spectroscopy of the Russian Academy of Sciences.

\section*{Declaration of competing interest}

The authors declare that they have no known competing financial interests or personal relationships that could have appeared to
influence the work reported in this paper.

\section*{Data availability}

All data underlying the results presented in this paper may be obtained from the authors upon reasonable request.





\printcredits


\bibliography{cas-refs}

\bio{}
\endbio


\end{document}